\documentclass[10pt,twocolumn]{article}
 
\usepackage{booktabs}
\usepackage{multirow}
\usepackage[margin=1in]{geometry}
\usepackage[T1]{fontenc}
\usepackage{lmodern}
\usepackage{graphicx}
\usepackage{amsmath}
\usepackage{booktabs}
\usepackage{array}
\usepackage{url}
\usepackage[numbers,sort&compress]{natbib}
\usepackage[hidelinks]{hyperref}
\usepackage{tabularx}
\usepackage{subcaption}
\usepackage{amsmath}

\usepackage{listings}
\usepackage{xcolor}

\title{InfantConnect: integrating developmental brain atlases \\ and electrophysiological network analysis}

\author{
Alessandro Crimi$^{1}$ \quad
Paweł Gryboś$^{1}$\\[4pt]
\small $^{1}$AGH University of Krakow, Faculty of Computer Science, Krakow, Poland\\
\small Correspondence: [alecrimi@agh.edu.pl]
}

\date{}

\begin{document}
\maketitle

\begin{abstract}
The brains of infants change considerably during the first two years.  
Hence, developmental neuroimaging requires computational tools that account for age-related anatomy while supporting the analysis of electrophysiological data. 
For instance,  cortical surfaces,   electrode montages, tractography atlases, and electroencephalography (EEG) analysis utilities are commonly distributed across separate resources and software environments. Here we  propose  an open Python framework that brings these components into a shared three-dimensional workspace and supplements interactive visualization with command-line tools for EEG quality control, preprocessing, source-related visualization, and effective-connectivity analysis. The command line version can also be invoked by a model context protocol server to be used by artificial intelligence  agents. 
The tool supports age-indexed cortical surfaces, infant scalp and skull templates, electrode locations, tractography, sensor-level connectomes, and directed estimates based on Granger causality and transfer entropy. We demonstrate the resource using representative anatomical visualizations and evaluate the effective-connectivity workflow on controlled toy networks with explicitly specified nodes and directed interactions. These tests provide targeted evidence about the connectivity workflow rather than a general validation of causal inference from clinical EEG. Performance benchmarks on low resolution electromagnetic tomography  and validation of effective connectivity are provided. 
\end{abstract}

\begin{figure*}[!h]
\centering
\includegraphics[width=\textwidth]{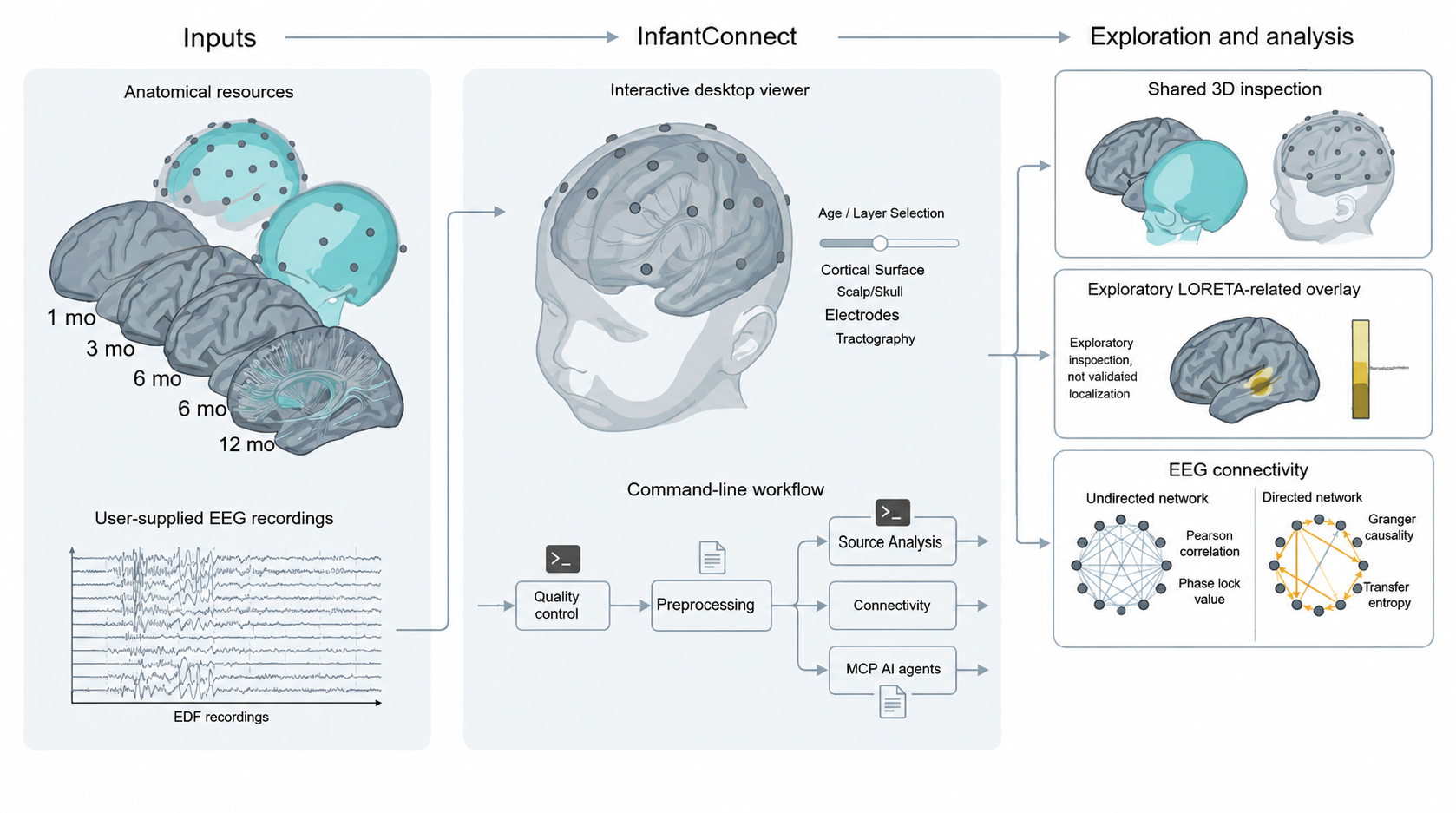}
\caption{ Overview of the InfantConnect architecture and reproducible workflow: desktop viewer, developmental cortical surfaces, infant scalp and skull templates, EEG electrode locations, tractography, EEG quality-control and preprocessing tools, source-related visualization, and effective-connectivity analysis.  Those elements can be used jointly via a graphical interface or individual command line scripts.}
\label{fig:pipeline}
\end{figure*}

\section*{Introduction}
The first years of life in humans are characterized by substantial changes in brain structure and function, as the brain continues to grow and change even after delivery \cite{zhang2019neonate}. 
Therefore, developmental neuroimaging  benefits from computational resources that adapt to age-related anatomical changes rather than assuming that adult templates are appropriate for all participants. 
Magnetic resonance imaging (MRI) and electroencephalography (EEG)  templates 
that adapt with age have been developed to support analyses across infancy and early childhood \cite{richards2016database,Li2015,zhang2019neonate,Rekik2016,Wang2019}. For EEG analysis,       electrode positions are relevant for proper developmental  anatomy evaluation and source localization, as well as the interpretation of cortical source estimates. Infant-specific head and surface templates can provide a more age-aware anatomical context for EEG visualization \cite{oreilly2021structural,Conte2022,Tsolisou2023}. 

An additional requirement is the ability to relate EEG-derived source estimates to anatomically interpretable brain regions. A source map displayed on a cortical surface provides spatial context, but without a corresponding parcellation it can be difficult to identify and consistently report the anatomical regions associated with the displayed activity. Age-associated cortical parcellations can provide this regional reference, allowing users to inspect source-related estimates in relation to named cortical regions and to select regions of interest (ROIs) for subsequent analysis \cite{mazurek2024explainable, mazurek2025graph}. This is particularly relevant in developmental neuroimaging, where the cortical geometry changes substantially with age, and adult anatomical references may not be directly appropriate.

Some of those tools have already been developed, though they are spread across different platforms and repositories, and they vary in user-friendliness.  In practice, researchers may need to locate anatomical templates, prepare surface files, manage electrode coordinates, inspect EEG recordings, compute connectivity, and possibly inspect tractography using different tools. In addition, these tools lack integration with agentic frameworks, which is a growing interest. This fragmentation makes it harder to inspect the spatial relationships between representations and increases the effort needed to reproduce a workflow. It also limits accessibility for researchers and students who need an integrated environment for exploration and method development. Lastly, an aspect often neglected by infant neurotools is the use of effective brain connectivity to study causal relationships and further biomarkers in premature infants or other clinical cases \cite{hendrikx2020nonlinear}.  Beyond spontaneous electrophysiological activity, transcranial magnetic stimulation combined with electroencephalography (TMS-EEG) can be a relevant novel tool to evaluate perinatal brain injuries \cite{collins2025longitudinal,salehinejad2024safety}. In this setting, TMS provides a controlled cortical perturbation which can be defined as effective connectivity.

We propose a framework that addresses this workflow problem by combining developmental cortical surfaces, infant head and scalp geometry, EEG electrode positions, an age-associated cortical parcellation, tractography, and EEG-derived visualizations in a desktop application. The parcellation provides an anatomical ROI reference that can be inspected alongside cortical surfaces and LORETA-related source maps. We combine these in a seamless 
visualization tool, as well as in command line scripts to allow automated analysis and pre-processing of large datasets.  Hence, the current repository also provides command-line workflows for EEG quality checks and preprocessing, connectome export, Granger-causality analysis, transfer-entropy analysis, and low-resolution electromagnetic tomography (LORETA)-related visualization. The project integrates existing third-party anatomical resources rather than introducing a new atlas or claiming that visual alignment alone establishes anatomical correspondence. An overall view of the framework is depicted in Figure \ref{fig:pipeline}. 

InfantConnect includes utilities for EEG file quality checks and preprocessing in addition to the graphical workflow. EDF recordings can be assessed using a quality-reporting command-line tool that writes text and/or JSON reports. A separate preprocessing utility supports configuration-driven processing and command-line overrides for selected settings. These interfaces allow routine checks and preparation steps to be run independently of the GUI, supporting batch processing and the recording of analysis settings.

The preprocessing pipeline can be configured graphically or through a configuration file. The default settings include a 50~Hz notch filter followed by zero-phase band-pass filtering between 0.5 and 40~Hz, and common-average re-referencing. No cropping or resampling is performed by default, preserving the original recording duration and sampling frequency. These are software defaults rather than universal recommendations; the settings used for a particular analysis should be reported from the configuration actually applied.

The repository also provides command-line entry points for exporting Granger-causality and transfer-entropy connectomes as CSV files and for generating LORETA-related outputs for specified ages and surface configurations. These tools allow parts of InfantConnect to be integrated into scripted workflows, while the desktop application supports interactive spatial inspection. Exact preprocessing choices, channel labels, references, filtering, resampling, and rejected segments should be recorded for each analysis and should not be inferred from a screenshot alone.

\begin{lstlisting}[]
python edf_preprocess_cli.py --write-default-config > preproc.json

python edf_preprocess_cli.py rec.edf -c preproc.json --set lowpass=45 --set resample=true --set resample_hz=125

python edf_granger_cli.py rec.edf -o out/ --lag 5 --edge-list

python edf_transfer_entropy_cli.py rec.edf -o out/ --bins 4 --lag 1

python edf_loreta_cli.py rec.edf --age 12 -o out/
\end{lstlisting}

The contribution is therefore an open computational resource spanning data preparation, interactive visualization, and selected EEG analyses. We describe its components and reproducible use, and report targeted tests of effective connectivity using synthetic networks with known directed structure and other data. We distinguish these tests from the validation of source localization, anatomical registration, or clinical inference. The aim is to make the capabilities, dependencies, and limitations of the resource explicit so that it can be reused and extended by the developmental neuroscience community.

\begin{figure*}[!]
  \centering
  \begin{subfigure}{0.48\linewidth}
    \centering
    \includegraphics[width=\linewidth]{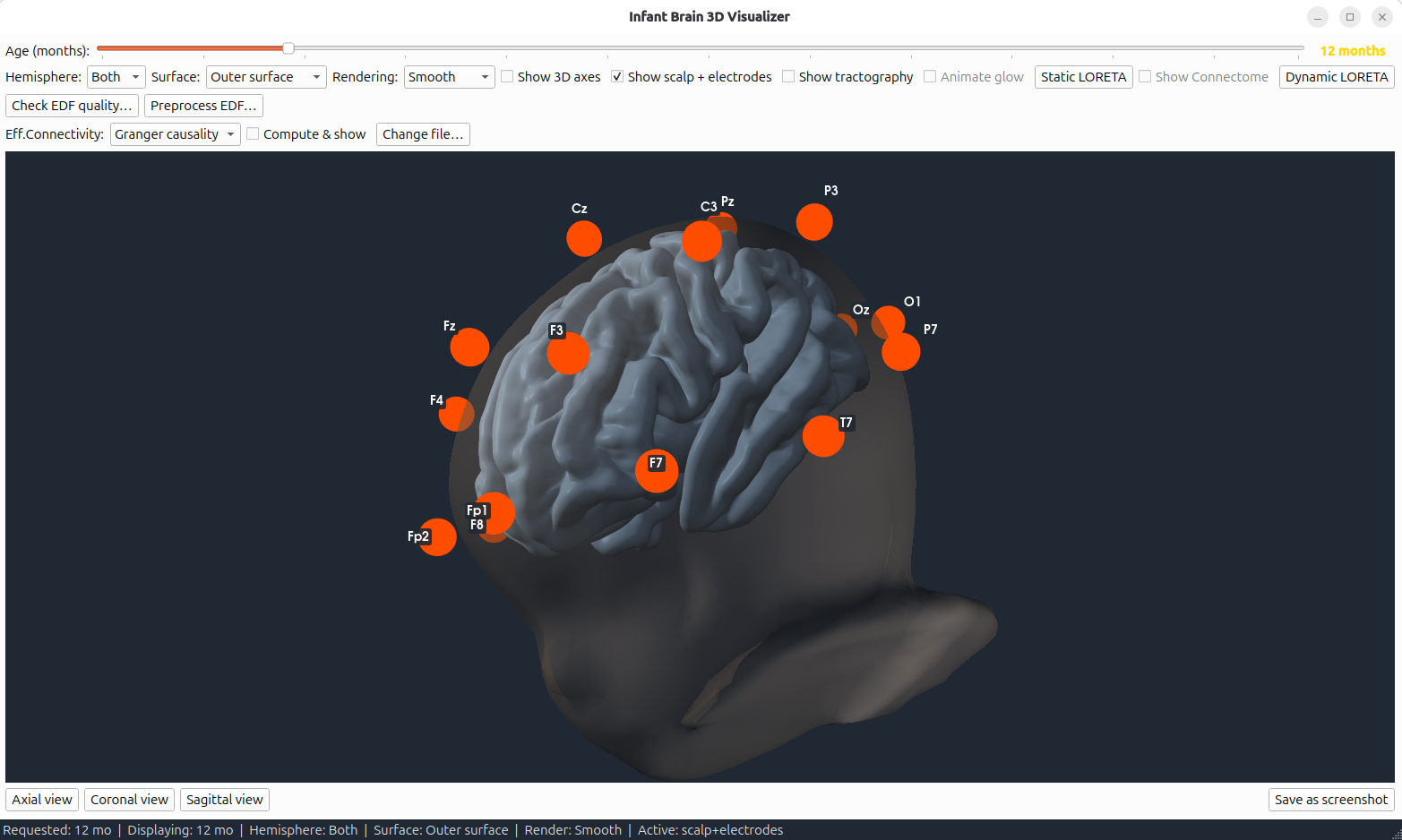}
    \caption{ }
    \label{fig:scr:a}
  \end{subfigure}\hfill
  \begin{subfigure}{0.48\linewidth}
    \centering
    \includegraphics[width=\linewidth]{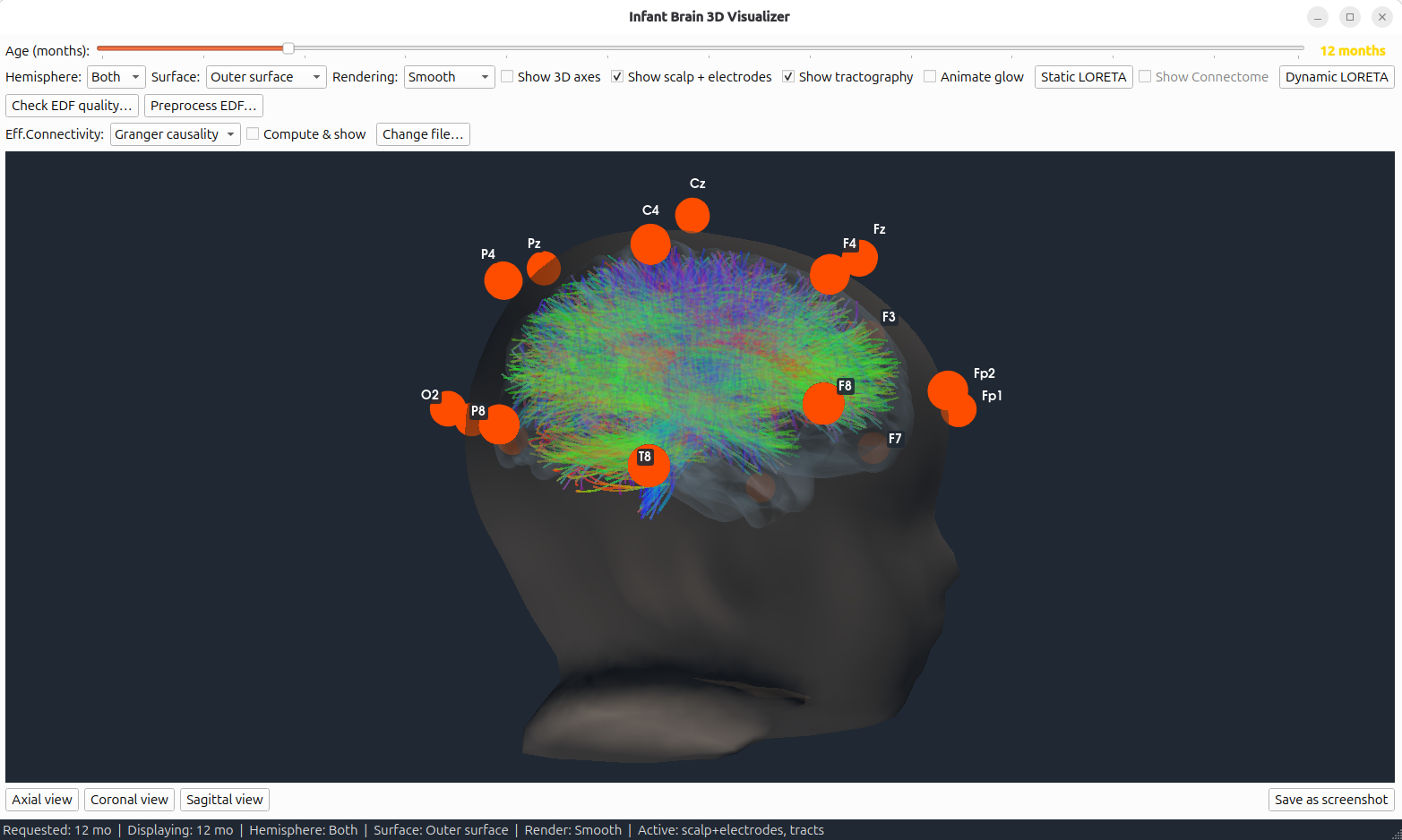}
    \caption{ }
    \label{fig:scr:b}
  \end{subfigure}

  \vspace{1ex}
  \begin{subfigure}{0.48\linewidth}
    \centering
    \includegraphics[width=\linewidth]{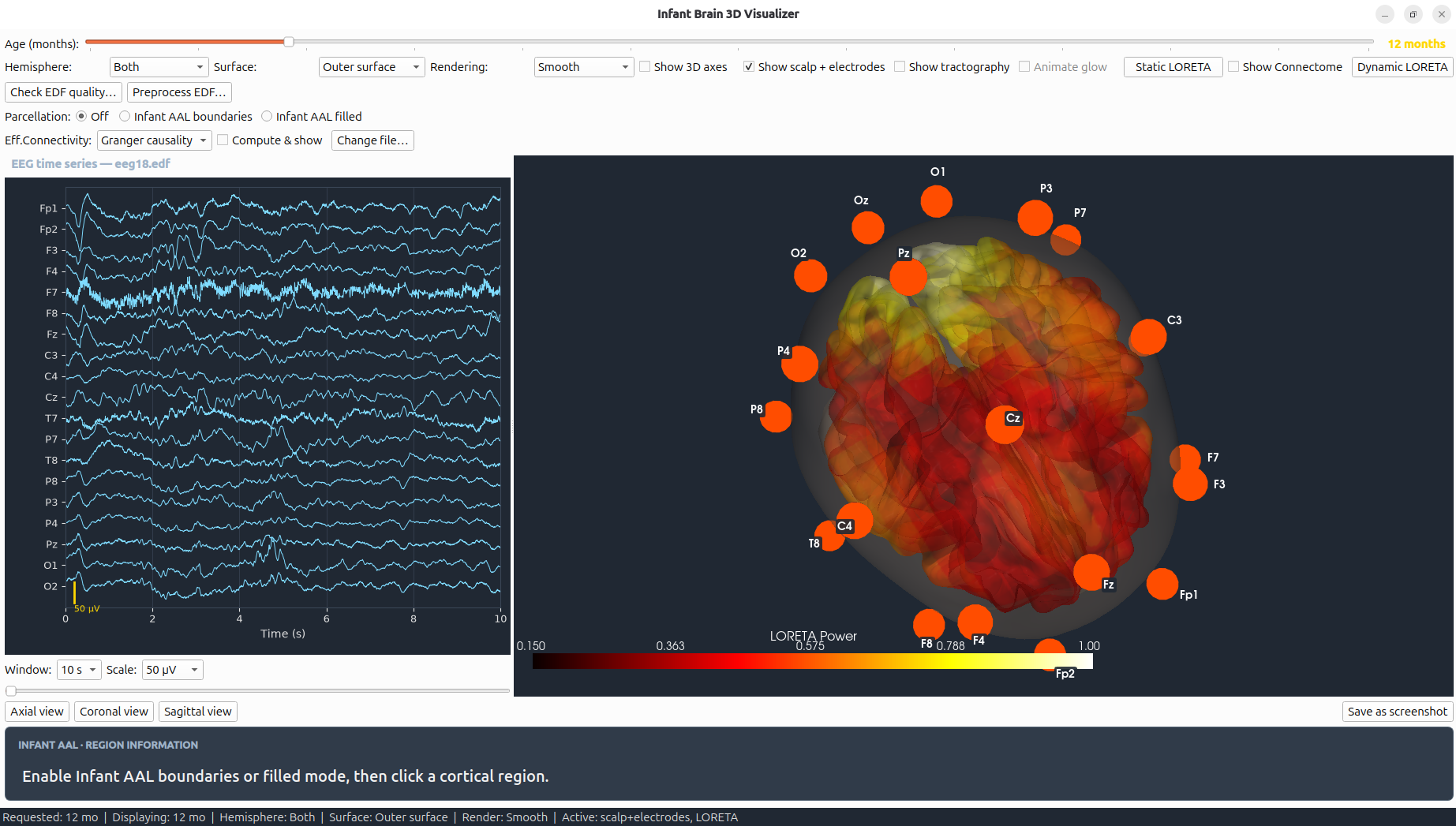}
    \caption{ }
    \label{fig:scr:c}
  \end{subfigure}\hfill
  \begin{subfigure}{0.48\linewidth}
    \centering
    \includegraphics[width=\linewidth]{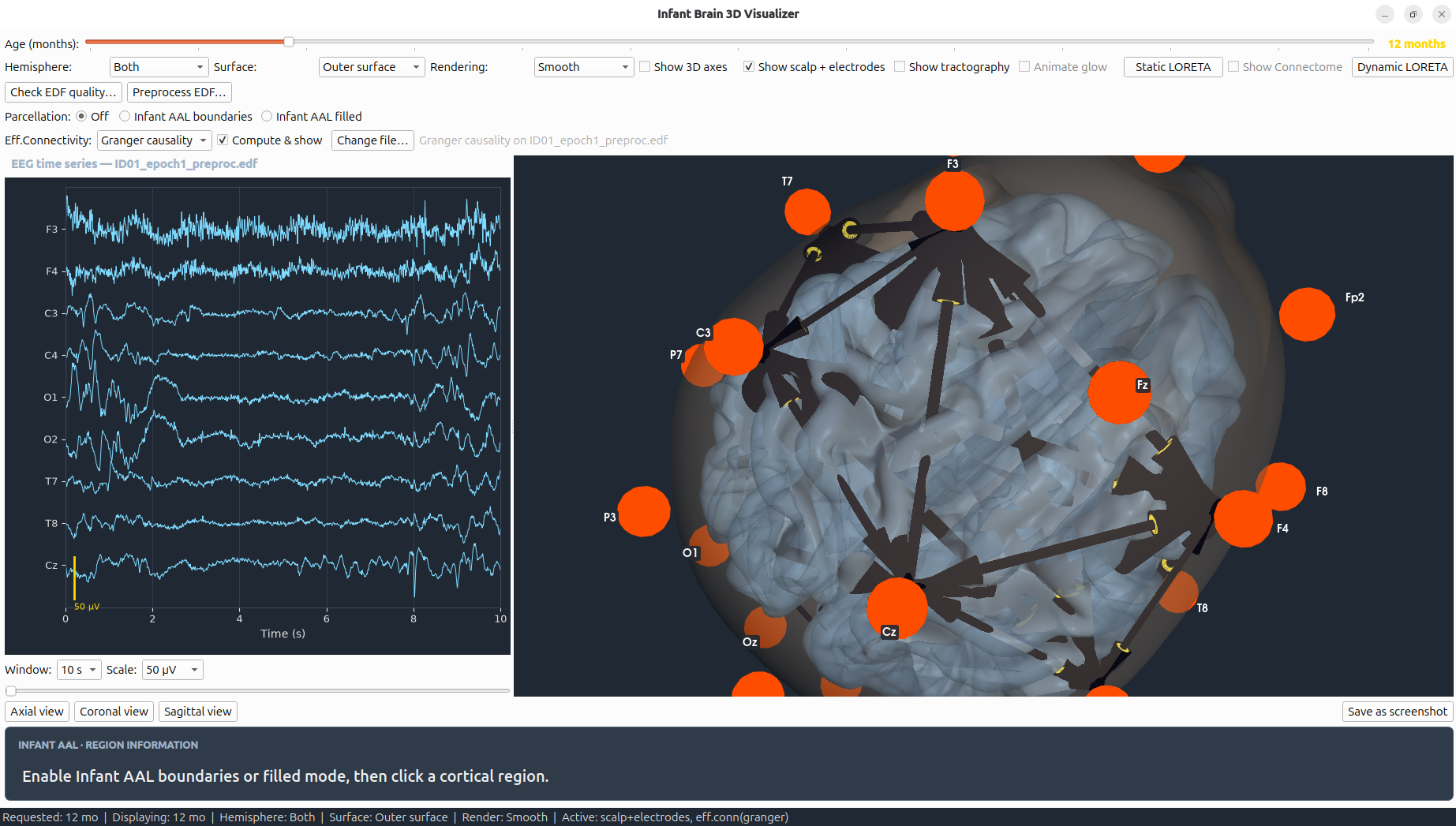}
    \caption{  }
    \label{fig:scr:d}
  \end{subfigure}

  \caption{Representative age-associated screenshots showing different
           operations: (a)~cortical surfaces and infant head/scalp geometry
           with EEG electrode positions, (b)~tractography in the shared 3D
           scene, (c)~mapped static LORETA coefficients, and (d)~effective
           connectivity estimated from a given \texttt{.edf} file.}
  \label{fig:screenshots}
\end{figure*}

\section*{Results}

\subsection*{An integrated environment for developmental brain visualization}

InfantConnect provides a desktop environment for inspecting developmental brain anatomy and EEG-related outputs in a shared three-dimensional scene. The application combines age-associated cortical surfaces, infant head and scalp geometry, EEG electrode positions, tractography, cortical parcellation, and EEG analysis outputs. Users can select an age, switch anatomical layers and surface representations, inspect the spatial relationships between resources, and save scenes for further inspection and communication.

The cortical surface resources are organized across developmental ages from 1 to 72 months, at three-month intervals where the corresponding prepared files are available. The age selector provides a convenient means of choosing an age-associated surface and related visualization resources. Availability and completeness depend on the files included in the local resource manifest. The age selector is therefore an organizational feature and does not imply that every age, surface representation, or anatomical structure has equivalent coverage or independent validation.

Figure~\ref{fig:screenshots} illustrates representative visualization scenarios, including developmental surface inspection, EEG electrode placement, LORETA-related maps, and directed connectivity. The shared scene allows these elements to be inspected together, while their anatomical correspondence remains dependent on the provenance and spatial preparation of the underlying resources.

\subsection*{Anatomical resources and resource quality control}

InfantConnect integrates the 4D UNC Neonatal and Infant Cortical Surface Atlases \cite{li2015uncinfant}, infant AAL parcellation \cite{shi2011infant}, infant head and skull templates associated with MNE-related work \cite{oreilly2021structural}, and the NABA800FC tractography atlas \cite{zhang2026cross}. Preparation and retrieval scripts support cortical-surface organization and tractography acquisition, while head, skull, and electrode/montage templates can be retrieved on demand and cached locally. These resources are kept separate from the application logic, allowing external assets to be updated or extended without redesigning the viewer.

A manifest-based quality-control procedure was used to inspect the prepared resource collection as also shown in Table~\ref{tab:qc}. All 110 successfully processed assets passed the automated integrity checks,
with no detected file or manifest errors. Maximum symmetric extent ratio relative to the median of 3.246 indicates that the largest observed extent was up to
324.6\% greater than the median extent across assets.
Such variation is due to the fact that sphere meshes are bigger that other types. Importantly, surface extents are reported in
the original input coordinate units and are not interpreted as a
direct measure of anatomical accuracy or registration quality.
To demonstrate the validity of the workflow we tested the command line steps on the whole Helsinki dataset (details about the dataset and the used machine are given in the Material section), and we provide the runtime and scale for each individual step in Table \ref{runtime}.

\begin{table*}[t]
\centering
\small
\renewcommand{\arraystretch}{1.15}
\caption{Asset integrity and geometric consistency checks for the
InfantConnect resource manifest. Surface extents are reported in the
input coordinate units of the corresponding atlas meshes and represent
medians across all successfully processed assets. The maximum extent
ratio is calculated relative to the median extent for the corresponding
coordinate dimension.}
\label{tab:qc}
\begin{tabularx}{\textwidth}{@{}X r@{}}
\toprule
Quality control metric & Result \\
\midrule
Meshes in population & 110 \\
Meshes with valid geometry & 110 \\
Meshes with detected errors & 0 \\
Median surface extent, $X$ & 61.62 \\
Median surface extent, $Y$ & 149.24 \\
Median surface extent, $Z$ & 99.94 \\
Maximum symmetric extent ratio relative to the median & 3.246 \\
\bottomrule
\end{tabularx}
\end{table*}

\begin{table}[t]
\centering
\caption{Runtime evaluation of the InfantConnect EEG processing
and analysis workflows. Values are reported as mean $\pm$ standard
deviation across recordings, in seconds. Here we also provide a comparison 
between our implementation of Granger and the Statsmodel implementation}
\label{runtime}
\begin{tabular}{lc}
\hline
\textbf{Operation} & \textbf{Runtime (s)} \\
\hline
EEG quality assessment       & $1.47 \pm 0.27$ \\
EEG preprocessing            & $1.68 \pm 0.28$ \\
LORETA source localization   & $5.58 \pm 0.58$ \\
Granger causality (selfcontained)           & $22.48 \pm 3.66$ \\
Granger causality  (Statsmodel)          & $45.56 \pm 1.57$ \\
Transfer entropy             & $1.99 \pm 0.45$ \\
\hline
\end{tabular}
\end{table}

\subsection*{Age-associated cortical parcellation and ROI coverage}

InfantConnect provides a cortical parcellation based on infant AAL anatomical labels. While the developmental cortical surfaces cover approximately 1--72 months at three-month intervals, the volumetric infant AAL labels available in the current implementation correspond to three developmental time points: neonatal (0 months), 12 months, and 24 months. These volumetric labels are projected onto the available cortical surfaces to generate hemisphere-specific surface label arrays.

Consequently, the cortical surface and parcellation resources have different temporal resolutions. The cortical geometry can be selected at approximately three-month intervals, whereas the anatomical labels originate from only three volumetric atlas time points. Parcellations displayed at intermediate surface ages therefore represent projections of the available AAL label maps rather than independently defined age-specific parcellations.

Across the 22 generated hemisphere-specific surface label arrays, nonzero-label coverage ranged from 76.1\% to 91.8\%. Each hemisphere contained 37--40 represented regions, corresponding to approximately 125,000--150,000 labelled vertices. All 22 arrays contained nonzero labels, indicating successful generation of the surface parcellation files.

\begin{figure}[t]
    \centering
    \includegraphics[
        width=\columnwidth,
        keepaspectratio
    ]{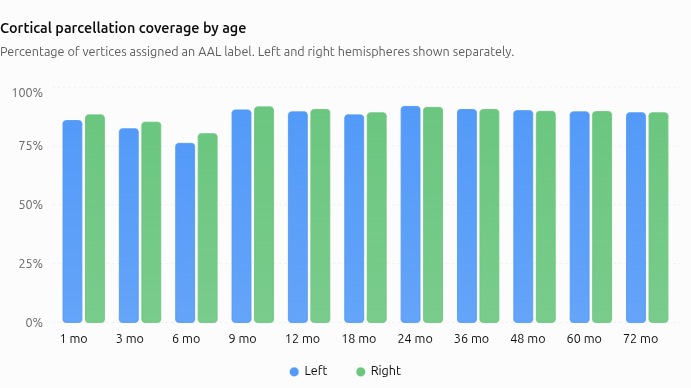}
    \caption{     Coverage and regional representation of the infant AAL cortical parcellation.     The 22 generated hemisphere-specific surface label arrays were evaluated for nonzero vertex coverage and the number of distinct anatomical regions represented, and are here depicted per hemisphere and month. 
    }
    \label{fig:infant_aal_coverage}
\end{figure}

Figure~\ref{fig:infant_aal_coverage} summarizes the coverage and regional representation across the generated arrays. These measurements quantify the proportion of cortical mesh vertices assigned a nonzero atlas label. They provide a technical assessment of the projection and its coverage across the available surfaces, but do not independently establish anatomical accuracy or age-specific correspondence of the ROIs.

The parcellation can be displayed as filled cortical regions or regional boundaries. Selecting a region in the graphical interface returns its anatomical acronym, full name, and hemisphere. This ROI identification is available in the same visualization environment as the LORETA-related maps, allowing users to inspect source-related values in the context of named anatomical regions.

\subsection*{LORETA-related visualization and ROI inspection}

\begin{figure*}[!h]
\centering
\includegraphics[width=\textwidth]{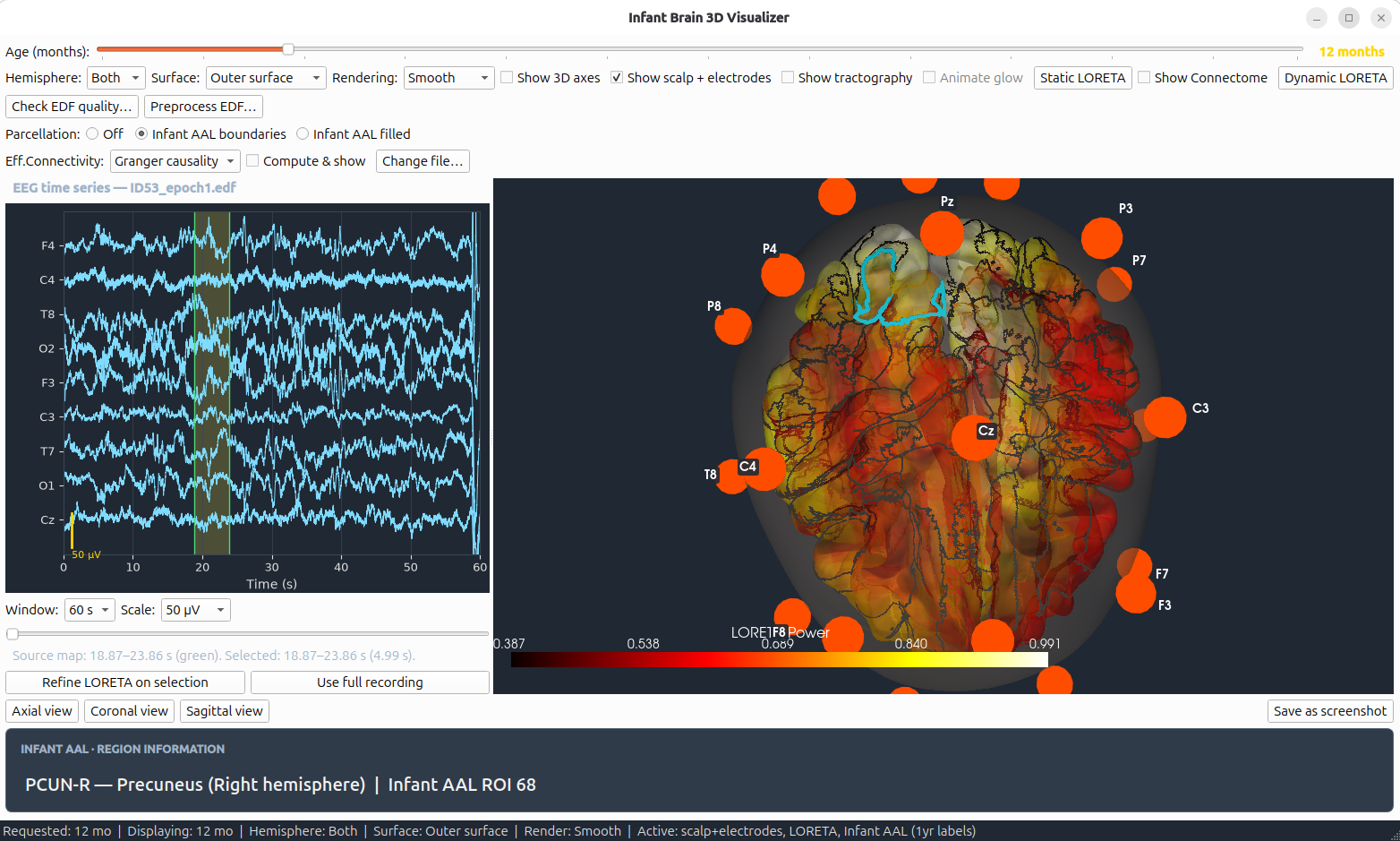}
\caption{
LORETA-related values displayed on the infant AAL cortical parcellation for a subject from the Helsinki neonatal EEG dataset \cite{stevenson2018neonatal}. The right precuneus is highlighted as a region with a relatively high displayed value. The remaining parcellation provides anatomical context and supports identification of the selected ROI. This example demonstrates the visualization and ROI-inspection workflow; it does not independently validate source localization.
}
\label{fig:loretaroi}
\end{figure*}

InfantConnect provides a workflow for displaying LORETA-related scores or source estimates on a selected developmental cortical surface. The cortical parcellation provides an additional anatomical reference for inspecting these maps. Users can display regional boundaries or filled anatomical regions and select an ROI to retrieve its anatomical identity, including its acronym, full name, and hemisphere.
Simultaneously, the graphic interface shows the EEG time series, which might further help to understand the LORETA localization. This can be a whole time series or a section. 
The tools can highlight  thicker time series (as F7 in Figure \ref{fig:scr:c}) because their signals contain much faster activity than the others. At 10s per window, a channel with high-frequency content swings up and down many times per pixel. The line then fills in as a dense band instead of a thin curve. This might be a sign of muscle artifacts, poor electrode contact, or high impedance, and further preprocessing should be carried out. The same approach can be used to inspect the location of a seizure. \textbf{Indeed, the tool allows the selection of suspicious sections of the time series representing seizures or burst  suppression, which can be further inspected on the 3D model visualization as shown in Figure  \ref{fig:loretaroi}}. 

Figure~\ref{fig:loretaroi} shows an example from a subject in the Helsinki neonatal EEG dataset \cite{stevenson2018neonatal}. In this example, the right precuneus is highlighted as a region with a relatively high displayed value, while the remaining cortical parcellation provides anatomical context. The figure illustrates the combined visualization of a LORETA-related map and atlas-defined ROIs, rather than an independently validated localization result.

The ROI interface allows the user to move from a continuous source-related map to the identification of named cortical regions. This can support regional inspection and reporting, provided that the source-estimation procedure and the mapping between source space and displayed surface are documented. Displaying a source-related value on a developmental surface does not itself demonstrate accurate source localization. The interpretation depends on the EEG montage and reference, data quality, forward model, source-estimation implementation, and spatial correspondence between the source space and the displayed surface \cite{Hamalainen1994,Dale2000,PascualMarqui2002,Conte2022}.

\subsection*{Effective-connectivity workflow and controlled benchmark}

InfantConnect supports Granger causality and transfer entropy for estimating directed statistical relationships between EEG channels. The software can render estimated directed relationships as edges between nodes and export connectivity results for further analysis. Example settings include a Granger lag of five and transfer entropy with four bins and a lag of one. These settings are implementation examples rather than universal choices for infant EEG.

\begin{figure}[t]
\centering
\includegraphics[width=\columnwidth]{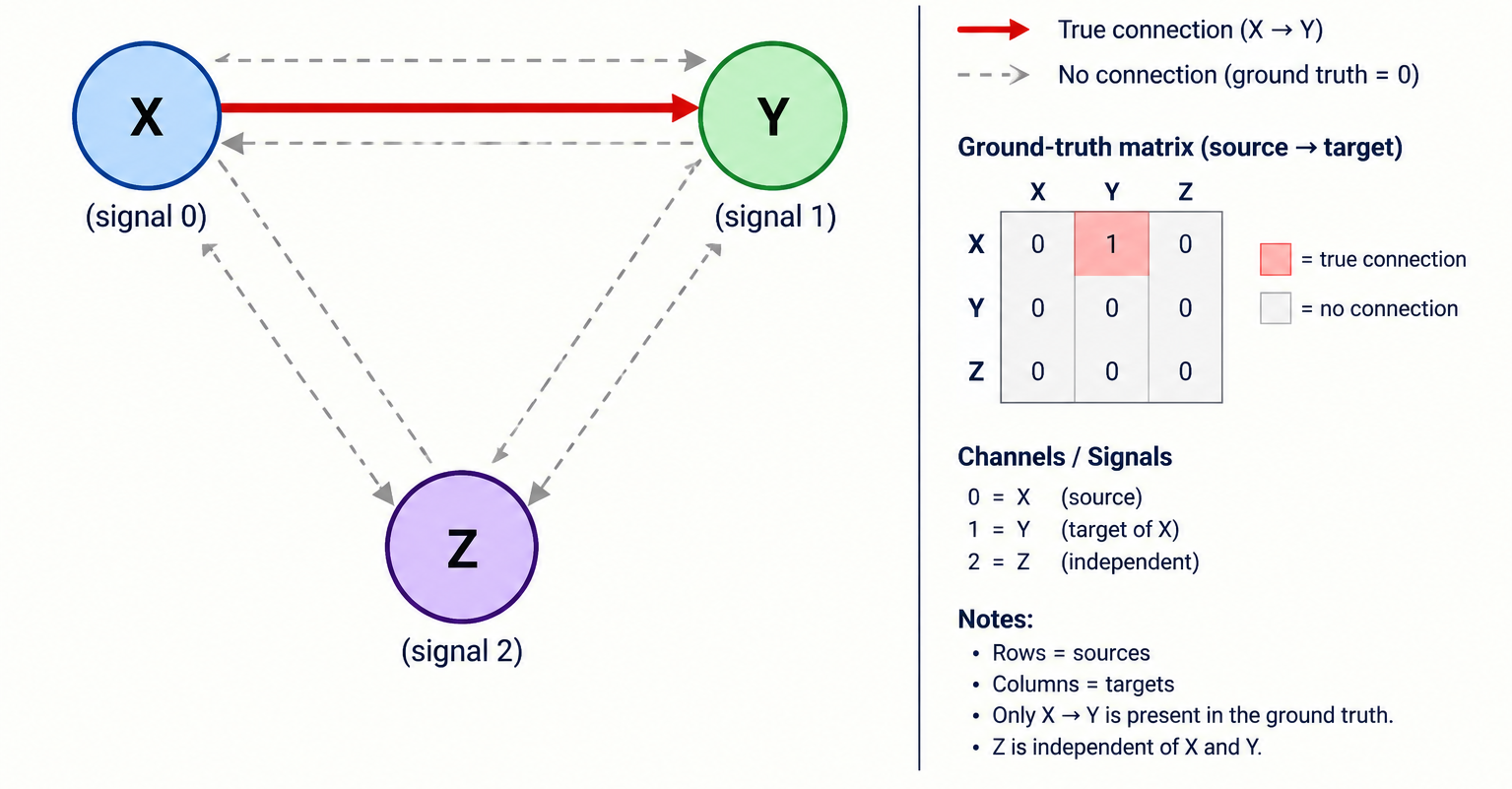}
\caption{
 Effective-connectivity estimation on a synthetic network with known directed structure.  The ground-truth network specifies a single directed interaction, $X\rightarrow Y$, among three simulated signals, the time series evolving in the specific nodes can be related by linear or nonlinear relationships.  
}
\label{fig:causal}
\end{figure}

The effective-connectivity workflow was evaluated using two complementary test cases. First, synthetic time series were generated from a network with explicitly defined directed interactions. This controlled setting provides a known reference structure against which estimated networks can be compared. The ground-truth network contains three nodes and a single directed connection from $X$ to $Y$ (Figure~\ref{fig:causal}). The synthetic data generation procedure, ground-truth network, and evaluation protocol are described in detail in the Synthetic effective-connectivity benchmark subsection of Methods. In our experiments, both Granger causality and transfer entropy recovered the known interaction in the reported experiment in Table \ref{tab:ec_linear_nonlinear}, though, as expected, Granger causality degraded if non-linear relationships were used in the model, which are typical  in more complex or realistic systems  \cite{ciezobka2025end}.

\begin{table*}[t]
\centering
\small
\renewcommand{\arraystretch}{1.2}
\caption{Performance of Granger causality and transfer entropy (TE) 
for effective-connectivity estimation in linear and nonlinear 
synthetic systems, mean values of 2000 runs.}
\label{tab:ec_linear_nonlinear}
\begin{tabular}{llccccc}
\toprule
System & Method & AUC & Average Precision & F1 & Precision & Recall \\
\midrule
\multirow{2}{*}{Linear}
& Granger & 0.9220 & 0.8267 & 0.6167 & 0.5750 & 0.7000 \\
& TE      & 0.8620 & 0.7553 & 0.4650 & 0.4233 & 0.5500 \\
\midrule
\multirow{2}{*}{Nonlinear}
& Granger & 0.616 & 0.260 & 0.333 & 0.279 & 0.414 \\
& TE      & 0.974 & 0.908 & 0.811 & 0.795 & 0.829 \\
\bottomrule
\end{tabular}
\end{table*}

Second, the workflow was applied to the Helsinki Dataset of neonatal EEG recordings with seizure annotations \cite{stevenson2018neonatal}. This example demonstrates that EEG data from an infant cohort can be passed through the analysis and visualization workflow. Because the dataset does not provide a ground-truth effective-connectivity network, the real-recording example is a workflow demonstration and does not support quantitative evaluation of connectivity-estimation accuracy.

\section*{Discussion}

InfantConnect brings developmental cortical surfaces, an infant AAL-based cortical parcellation, infant head and scalp templates, EEG electrode locations, tractography, and EEG analysis outputs into a shared computational environment. The resource combines anatomical visualization with practical EEG workflows, allowing users to inspect developmental anatomy, prepare recordings, estimate directed statistical relationships, and visualize source-related outputs within a common interface. Its principal contribution is the integration of reusable anatomical resources and analysis tools in a very friendly and automated manner, rather than the introduction of a new EEG inverse method or a new effective-connectivity estimator.

A central feature of the framework is the relationship between the cortical parcellation and LORETA-related visualization. The infant AAL-based surface labels provide a means of identifying named cortical ROIs in the context of source-related maps. The coverage analysis shows that the generated hemisphere-specific arrays contain nonzero labels across the evaluated surfaces, with coverage ranging from 76.1\% to 91.8\%. This provides a technical measure of how much of each cortical mesh is assigned an atlas label and helps characterize the generated parcellation resources. However, coverage is not equivalent to anatomical accuracy: a vertex can receive a nonzero label without that label being anatomically correct at that location.   An important distinction concerns the different developmental resolutions of the anatomical resources. Although cortical surfaces are available at approximately three-month intervals across the first six years of life, the infant AAL volumetric labels currently cover only neonatal, 12-month, and 24-month time points. Consequently, the ROI maps displayed on intermediate-age cortical surfaces are projections of these available atlas volumes rather than independently age-specific parcellations. This distinction is important when interpreting regional anatomy and comparing ROI measurements across developmental ages. Future extensions could incorporate additional age-specific volumetric atlases or validated longitudinal label-propagation procedures to improve the temporal correspondence between cortical geometry and anatomical parcellation.

The LORETA example from the Helsinki dataset illustrates how source-related values can be displayed together with the infant AAL parcellation and how a user can inspect a region such as the right precuneus. This is useful as a demonstration of the workflow and as a means of organizing regional inspection and reporting. It should not be interpreted as independent validation of source localization. 

The framework also supports dynamic LORETA-related time series that can be saved in HDF5 files. This output may facilitate downstream analysis of regional or source-related signals, but the interpretation of such time series depends on the source-estimation configuration and the anatomical mapping used. The saved outputs should therefore be accompanied by the relevant processing and model parameters.

The effective-connectivity benchmark provides a controlled test of the software workflow. In the three-node autoregressive system, the only specified interaction was $X\rightarrow Y$, and both Granger causality and transfer entropy achieved acceptable estimates, though they degenerate in the case of non-linear causal relationships. This difference is consistent with the ability of transfer entropy to capture nonlinear statistical dependencies that are not adequately represented by a linear Granger model  \cite{ciezobka2025end, schreiber2000}. 
The Helsinki dataset example complements the simulation by demonstrating use with real recordings, but it cannot provide a quantitative estimate of connectivity accuracy because the underlying directed network is unknown.

The use of Granger causality and transfer entropy also requires careful interpretation. Both methods estimate directed statistical dependencies, but neither establishes causal interactions in the intervention-based sense. Their estimates can be affected by common drivers, non-stationarity, noise, filtering, reference choice, signal length, and parameter selection. A directed edge in a connectome should therefore be understood as an estimator output under specified analysis conditions, not as proof of a physiological causal connection. We observed that our implementation of Granger causality using optimized LAPACK routines performs better on the used hardware compared to the implementation used with Statsmodel. 

Several limitations remain. First, the anatomical components depend on third-party resources, and their completeness and provenance must be verified from the files used in a particular release. Second, the current geometry workflow may use scaling and translation to place heterogeneous assets in a shared scene.  Third, the parcellation coverage values describe label assignment on the cortical meshes, not the anatomical accuracy of the assigned labels. Finally, the synthetic effective-connectivity benchmark tests a restricted system and does not establish performance on more complex or clinically heterogeneous EEG recordings since ground truth on infant dataset are not available. 
 
Because the pipeline operates on \texttt{.edf} recordings, synthetic EEG can be generated and passed to it directly, with the choice of source model depending on the question. In every case, source signals are placed on an age-matched infant cortical surface \cite{li2015atlases}, projected to the scalp through a forward model built on infant head templates \cite{oreilly2021templates} using MNE-Python \cite{gramfort2013mne},   and exported as an \texttt{.edf} file. To benchmark source localization \cite{pascualmarqui1994loreta} and the connectivity analyses of \texttt{Infantconnect}  against a known ground truth, prescribed time series or neural mass models are the simplest choice: lumped-parameter models of coupled cortical columns can reproduce EEG activity and evoked potentials \cite{jansen1995neuralmass}, and in the generalized model of David and Friston \cite{david2003neuralmass} both coupling and propagation delay shape the simulated spectrum, so the interaction structure between regions is set explicitly. Networks of such mass models can be simulated over a structural connectome with The Virtual Brain \cite{sanzleon2013tvb}, a platform designed to generate macroscopic signals; for infants, this would require an age-appropriate connectivity and delay structure, for instance derived from tractography. When the question concerns cellular mechanisms, detailed multicompartment networks can be specified and simulated with NetPyNE \cite{durabernal2019netpyne}, whose built-in outputs include local field potentials, so an additional dipole and lead-field step would be needed to obtain scalp EEG. In the reverse direction, a source time course estimated from a real or synthetic recording can be summarized as a regional current-dipole waveform and used as the target for the Human Neocortical Neurosolver \cite{neymotin2020hnn,jas2023hnncore}, whose template circuit derives from a laminar model of somatosensory cortex constrained by human MEG \cite{jones2007tactile}, to propose candidate cellular and network mechanisms. Because that model was not developed for the infant cortex, applying it to infants would require re-parameterizing cell morphology, synaptic properties, and cortical geometry for the relevant age, and the resulting fits should be read as hypotheses rather than measurements.   Future work should indeed include more integration between simulations and the decomposition of EEG to and from microscopic biophysical models.

In summary, InfantConnect is an open computational resource for developmental brain visualization and EEG analysis. Its combination of cortical surfaces, infant AAL-based ROI identification, EEG preprocessing, connectivity estimation, LORETA-related visualization, and command-line workflows supports exploratory analysis and reproducible reporting. The controlled connectivity benchmark and the real-recording workflow example provide complementary evidence of functionality, while the limits of the parcellation coverage, spatial overlays, and source-related maps remain explicit. This combination of reusable software, documented data provenance, scripted workflows, and appropriately scoped evaluation provides the basis for presenting InfantConnect as a computational resource.

\section*{Methods}

\subsection*{Helsinki Dataset and experimental settings}
 The EEG workflow was demonstrated using the publicly available
Helsinki neonatal EEG dataset \cite{stevenson2018neonatal}, which
comprises recordings from 79 term neonates admitted to the Neonatal
Intensive Care Unit at Helsinki University Hospital. The recordings
were acquired using 19 EEG electrodes positioned according to the
international 10--20 system, at a sampling frequency of 256~Hz.
The median recording duration was 74~min. 

All experiments were performed on a laptop running Ubuntu,
equipped with an Intel\textsuperscript{\textregistered}
Core\textsuperscript{\texttrademark} 5 120U processor
(10 physical cores and 12 logical CPUs) and 15~GiB of RAM, no GPU was present.

\subsection*{Software architecture and installation}

InfantConnect is implemented in Python as a desktop application using several libraries either via user interface or command line. PySide6 provides the graphical user interface, while PyVista and the Visualization Toolkit (VTK) provide three-dimensional rendering. PyVista offers a Python interface to VTK-based scientific visualization \cite{Schroeder2006,Sullivan2019}. EEG processing uses the scientific Python ecosystem and MNE-related tools. The effective connectivity estimates are written as self-contained code rather than using widely used libraries such as Statsmodels \cite{seabold2010statsmodels}. Although, we compared Statsmodels to our implementation. The repository recommends Python 3.10 or newer and provides a requirements file. Software and dependency versions are relevant to reproducibility because scientific Python results can depend on execution details and package versions \cite{Sandve2013,Wilson2014}.

The main application is launched from the repository root after preparing the cortical surfaces and retrieving the tractography atlas. Infant scalp, skull, and electrode/montage template files can be downloaded on demand and cached locally. The source code is available at \url{https://github.com/alecrimi/infantconnect}, and the archived resource is available at \url{https://zenodo.org/records/23860029}. The package can be installed using:

\begin{lstlisting}[]
pip install infant-brain-viewer
\end{lstlisting}

\begin{figure}[!h]
\centering
\includegraphics[width=\columnwidth]{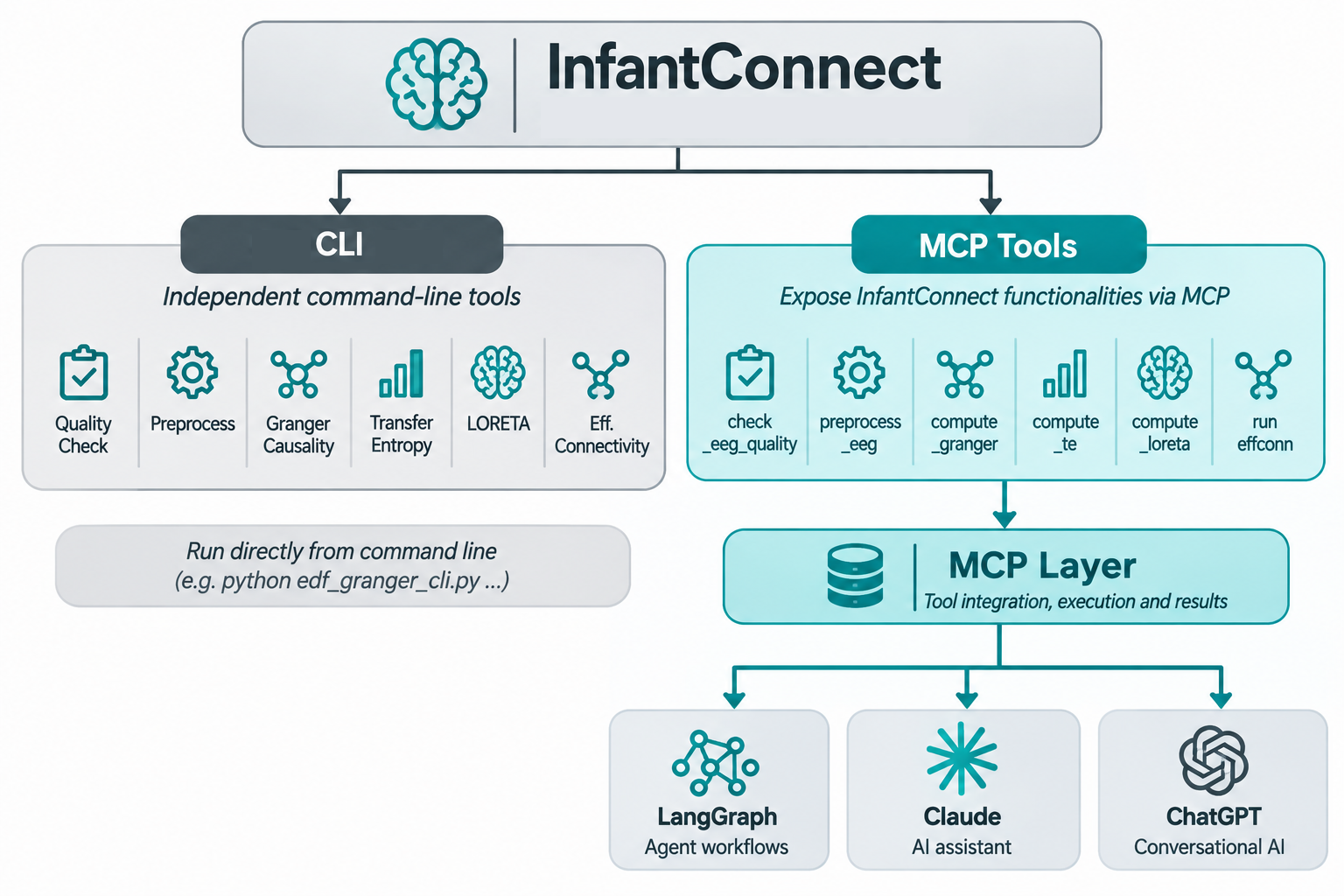}
\caption{ Overview of the possible automated workflow, where instead of using the user interface, the tools are called either by command line or by artificial intelligence  (AI) agents via an MCP server.}
\label{MCP}
\end{figure}

To facilitate the integration of InfantConnect with large language model (LLM)-based assistants and autonomous research agents, the software additionally provides an optional Model Context Protocol (MCP) server. MCP is an open communication protocol that standardizes how external applications expose computational tools and data resources to AI systems. Through this interface, the framework functionality can be invoked by compatible AI clients without requiring each client to implement its own interface to the underlying EEG processing routines.

The MCP implementation is provided as an additional software layer within the \texttt{infant\_brain\_viewer} Python package. It uses the Python MCP SDK and exposes the existing CLI functionality through structured tool definitions. The MCP server does not reimplement the  algorithms; instead, it validates the input parameters, invokes the corresponding CLI scripts, and returns execution status, textual output, and references to generated files. This architecture preserves a single implementation of each computational method while supporting both conventional command-line execution and AI-mediated interaction as shown in Figure \ref{MCP}.


The MCP interface allows InfantConnect to be used by different AI-agent environments without changing the underlying computational implementation. Three representative integration scenarios have been considered:

\textit{LangGraph.} LangGraph is a framework for constructing stateful, tool-using AI agents. Using the LangChain MCP adapters, an agent can discover the tools exposed by the InfantConnect server and incorporate them into a reasoning and execution workflow. For example, an agent can be instructed to assess the quality of an EEG recording, preprocess the data, calculate Granger causality and transfer entropy, and organize the resulting files.  

\textit{Claude.} Claude Desktop supports local MCP servers, allowing InfantConnect to be registered as an external tool provider. The server can be configured through the Claude Desktop configuration file. For example:

\begin{lstlisting}
{
  "mcpServers": {
    "infantconnect": {
      "command": "/path/to/python",
      "args": [
        "-m",
        "infant_brain_viewer.mcp_server"
      ],
      "env": {
        "INFANTCONNECT_MCP_TRANSPORT": "stdio"
      }
    }
  }
}
\end{lstlisting}

After restarting the application, Claude can discover the available InfantConnect tools and invoke them in response to natural-language instructions. For example, a user can request an EEG quality assessment or ask for connectivity analysis using specified parameters. The computational operations are executed by the local InfantConnect installation rather than by the language model itself.

\textit{ChatGPT.} InfantConnect can also be integrated with ChatGPT through a remotely accessible MCP server. In this configuration, the server is deployed using the Streamable HTTP transport and exposed through an HTTPS endpoint. A user can register the endpoint as a custom MCP application in a compatible ChatGPT environment. Once connected, the tools are made available to the assistant and can be invoked through natural-language requests.

For example, a user can request: ``Analyse this infant EEG recording, assess signal quality, calculate Granger causality and transfer entropy, and return the generated connectivity matrices.'' ChatGPT can select and invoke the corresponding tools, pass the required parameters, and present the returned results to the user. Unlike the local STDIO configuration, this deployment requires a remotely accessible server and appropriate authentication and filesystem access controls. Availability of custom MCP integrations and write-capable tools depends on the ChatGPT subscription and workspace configuration.

These integrations illustrate how InfantConnect can serve both as a conventional standalone scientific software package and as a computational resource accessible to AI agents. Importantly, the MCP layer is optional and does not replace the existing desktop application or command-line interface. The same EEG processing scripts remain independently executable, while MCP provides an additional standardized mechanism for orchestrating their functionality through different AI platforms.

\subsection*{Anatomical resources and geometry}

The application reads an atlas manifest to identify available cortical surfaces and supports selection by age, hemisphere, and surface representation when the corresponding files are present. Preparation scripts generate or organize cortical surface files, and a separate script retrieves tractography resources. External infant head, skull, and electrode templates are retrieved from their upstream source. The exact data files and their versions should be recorded for each published analysis.


\subsection*{Cortical parcellation projection and coverage assessment}

The cortical parcellation was generated by projecting infant AAL volumetric labels onto developmental cortical surfaces. The source volumetric label maps provide anatomical region identifiers, which are sampled at locations corresponding to vertices of each cortical mesh. The projection is performed separately for the left and right hemispheres, resulting in hemisphere-specific arrays in which each vertex is associated with an anatomical label or assigned an unlabelled value when no valid atlas label is available.

The projection uses the coordinate transformation implemented by the preparation script to map cortical surface coordinates into the coordinate system of the volumetric atlas. The transformation, coordinate units, image orientation, and voxel-to-world mapping are part of the reproducibility requirements and should be reported alongside the atlas source and the age-specific volume selected for each surface. The source maps used in the current implementation are the neonatal, 1-year, and 2-year infant AAL volumes. The assignment of these maps to developmental surface ages is determined by the preparation procedure and should be documented explicitly.

For each generated surface label array, coverage was calculated as the proportion of cortical vertices assigned a nonzero anatomical label:

\begin{equation}
 \mathrm{Coverage}(\%) =
100 \times \frac{N_{\mathrm{labelled}}}{N_{\mathrm{vertices}}},   
\end{equation}
 
where $N_{\mathrm{labelled}}$ is the number of vertices with a nonzero label and $N_{\mathrm{vertices}}$ is the total number of vertices in the corresponding hemisphere mesh. The number of represented regions was calculated as the number of distinct nonzero anatomical labels in the array. The number of labeled vertices was also recorded for each hemisphere. These measures characterize the extent of label assignment and regional representation; they do not independently quantify anatomical accuracy or registration quality. Although coverage does not directly measure anatomical accuracy, it provides an indirect and readily interpretable indicator of atlas-to-surface correspondence: higher coverage indicates that a larger proportion of cortical vertices could be assigned an anatomical label under the applied coordinate transformation. Consequently, unusually low coverage may indicate incomplete spatial correspondence, mismatches in coordinate systems, or limitations in the atlas representation. Coverage should therefore be interpreted as a proxy for the spatial correspondence achieved by the projection procedure, rather than as a standalone validation of registration accuracy.

The resulting label arrays are loaded by the graphical application and used to generate filled parcellation surfaces or regional boundary meshes. Interactive ROI identification is performed by selecting a point on the cortical surface and retrieving the corresponding vertex label. The label is then mapped to its anatomical acronym, full region name, and hemisphere using the atlas metadata. Vertices without a valid anatomical label are not assigned a region name.

\subsection*{EEG input, quality control, and preprocessing}

EEG recordings are accepted in EDF format. The repository includes a quality-check utility that writes a report in text and/or JSON format and a configuration-driven preprocessing utility. The preprocessing interface supports command-line configuration and selected parameter overrides. The checks and preprocessing settings used in an analysis must be reported from the configuration actually applied. Users should verify channel names and montage compatibility before associating recordings with electrode positions.

The default preprocessing configuration includes a 50~Hz notch filter, followed by zero-phase   band-pass filtering between 0.5 and 40~Hz and common-average re-referencing. No cropping or resampling is performed by default. These settings are configurable and should not be assumed to have been applied to a recording unless confirmed by its processing configuration. Artifact handling, channel selection, rejected segments, reference, filtering, and any resampling should be reported for each analysis.

\subsection*{LORETA-related output and ROI display}

The repository provides a command-line workflow for generating LORETA-related output for a specified age and optional cortical surface representation. The resulting maps or scores are displayed in relation to the selected surface. LORETA and its standardized and exact variants are established families of EEG/MEG inverse methods, but their outputs depend on the forward model and assumptions used in the inverse solution \cite{PascualMarqui1994,PascualMarqui2002,PascualMarqui2011}.

The exact estimator configuration, source space, forward model, regularization, and mapping from source estimates to displayed geometry should be recorded from the version used for the manuscript. For ROI inspection, the cortical surface label array is used to associate selected vertices with the corresponding infant AAL anatomical label. The label metadata provides the ROI acronym, full name, and hemisphere. This association supports anatomical identification in the interface but does not establish the accuracy of the underlying source estimate or the spatial correspondence between the source space and the displayed surface.

In the absence of an independently validated infant-specific forward model and localization, the resulting display is treated as a performance benchmark rather than as a validated localization of epileptogenic or other clinical sources. The workflow can also generate dynamic LORETA-related time series for saving in HDF5 format; the source-estimation settings and anatomical mapping should be documented alongside these outputs.

\subsection*{Granger causality and transfer entropy}

The software exposes Granger-causality and transfer-entropy calculations through the effective-connectivity module and provides command-line interfaces that can export connectome results as CSV files. The documented example settings are a Granger lag parameter of five, transfer entropy with four bins, and a lag of one. These values are implementation settings, not universal recommendations. Moreover, they are treated independently from source localization, quality control and preprocessing, assuming the input signals are not raw, and processed with the other components of the framework. 
Granger causality tests whether the past of one signal improves the prediction of another signal beyond the information contained in the latter's own past \citep{Granger1969,Barnett2014,crimi2021structurally}. For two time series, $X_t$ and $Y_t$, the unrestricted autoregressive model for predicting $Y_t$ is

\[
Y_t = \sum_{k=1}^{p} a_k Y_{t-k}
      + \sum_{k=1}^{p} b_k X_{t-k}
      + \varepsilon_t,
\]

whereas the restricted model excludes the past of $X_t$:

\[
Y_t = \sum_{k=1}^{p} a'_k Y_{t-k}
      + \varepsilon'_t.
\]

The null hypothesis of no Granger causality from $X$ to $Y$ is $H_0:b_1=\cdots=b_p=0$. The magnitude of the predictive contribution can be quantified using the logarithmic ratio of residual variances estimated over the time series,

\[
\mathcal{F}_{X\rightarrow Y}
=
\ln\left(
\frac{\widehat{\sigma}^{2}_{R}}
     {\widehat{\sigma}^{2}_{U}}
\right),
\]

where $\widehat{\sigma}^{2}_{R}$ and $\widehat{\sigma}^{2}_{U}$ denote the residual variance estimates from the restricted and unrestricted models, respectively. Larger positive values indicate a greater reduction in prediction error when the past of $X$ is included. Statistical significance can be assessed using an appropriate hypothesis test, such as a nested-model F-test or likelihood-ratio test.

Transfer entropy provides an information-theoretic formulation of directed interactions \citep{schreiber2000}. It quantifies the conditional mutual information between the past of a source process and the future of a target process, given the target's own past. For discrete-valued processes, transfer entropy from $X$ to $Y$ is defined as

\[
T_{X\to Y}
=\sum_{\substack{y_{t+1}\\ y_t^{(l)},\,x_t^{(k)}}}
p(y_{t+1},y_t^{(l)},x_t^{(k)})
\log
\frac{p(y_{t+1}\mid y_t^{(l)},x_t^{(k)})}
     {p(y_{t+1}\mid y_t^{(l)})},
\]

where $x_t^{(k)}$ and $y_t^{(l)}$ denote the $k$- and $l$-length histories of the source and target, respectively. Equivalently,

\[
T_{X\rightarrow Y}
=
I(Y_{t+1};X_t^{(k)}\mid Y_t^{(l)}),
\]

where $I(\cdot;\cdot\mid\cdot)$ denotes conditional mutual information, measuring the information shared between two variables conditional on a third. Here, transfer entropy quantifies the additional information provided by the source history about the target's future, beyond that contained in the target's own history. A positive value indicates that knowledge of the source history reduces uncertainty about the target's future. Transfer entropy is expressed in bits when the logarithm is base two, or in nats when the natural logarithm is used.

Granger causality evaluates improvements in predictive performance, typically within an autoregressive framework, whereas transfer entropy quantifies directed statistical dependence without requiring a linear relationship. Neither method, in its bivariate implementation, establishes causal interactions in the intervention-based sense. Both can be affected by unobserved common drivers, non-stationarity, noise, filtering, signal length, and parameter selection \cite{Barnett2014,Bastos2016}. The provided framework has a self-contained implementation for those causality estimators, though.

\subsection*{Synthetic effective-connectivity benchmark}

The effective-connectivity workflow was evaluated using synthetic multivariate time series generated from a three-node stochastic autoregressive system. Each dataset consisted of three continuous signals, denoted by $X$, $Y$, and $Z$, with $N=5000$ samples. The signals were generated using both linear and non-linear causal relationships. For the linear case  the following first-order autoregressive equations were used:

\[
\begin{aligned}
X_t &= 0.5X_{t-1}+\varepsilon_t^{(X)},\\
Y_t &= 0.4Y_{t-1}+0.8X_{t-1}+\varepsilon_t^{(Y)},\\
Z_t &= 0.5Z_{t-1}+\varepsilon_t^{(Z)},
\end{aligned}
\]

where $\varepsilon_t^{(X)}$, $\varepsilon_t^{(Y)}$, and $\varepsilon_t^{(Z)}$ are mutually independent Gaussian white-noise processes, sampled independently at each time point from a standard normal distribution, i.e., $\varepsilon_t^{(i)}\sim\mathcal{N}(0,1)$. All signals were initialized at zero.
For the non-linear case, the used system was:
\[
\begin{aligned}
X_t &= 0.5\tanh(X_{t-1})+\epsilon_{X,t}\\
Y_t &= 0.4\tanh(Y_{t-1})+
0.8\tanh(X_{t-1})+\epsilon_{Y,t}\\
Z_t &= 0.5\tanh(Z_{t-1})+\epsilon_{Z,t}.
\end{aligned}
\] 

The system contains one known directed interaction: the previous state of $X$ influences the subsequent state of $Y$, with a coupling coefficient of 0.8. In contrast, $X$ and $Z$ evolve independently, and $Z$ has no direct or indirect connections with the other signals. The autoregressive coefficients introduce temporal dependencies within each signal, while the coupling term introduces a known inter-signal dependency with a one-sample delay.

The ground-truth directed connectivity matrix, with rows representing source signals and columns representing target signals, is

\[
\mathbf{A}_{\mathrm{GT}} =
\begin{bmatrix}
0 & 1 & 0\\
0 & 0 & 0\\
0 & 0 & 0
\end{bmatrix},
\]

where a value of one indicates the presence of a directed interaction. Thus, the only ground-truth connection is $X\rightarrow Y$.

The experiment was repeated using 100 independent random seeds to assess performance across stochastic realizations. Granger causality and transfer entropy were applied to the same simulated datasets for each seed, allowing a paired comparison between the methods.  We have previously investigate more realistic relationship in a functional magnetic resonance imaging study on adult stroke \cite{ciezobka2025end}. 

\subsection*{Real-recording workflow demonstration}

The EEG workflow was also applied to the Helsinki Dataset of neonatal EEG recordings with seizure annotations \cite{stevenson2018neonatal}. The dataset was used to demonstrate the processing and visualization workflow on real neonatal EEG recordings. Because the dataset does not provide a ground-truth directed connectivity network, no quantitative evaluation of effective-connectivity accuracy was performed on these recordings. The real-data example is therefore treated as a workflow demonstration rather than a validation of the inferred connectivity.

\subsection*{Tractography display}

Tractography is displayed using the NABA800FC fibre-cluster atlas \cite{zhang2026cross}. Streamlines are subsampled to 2,000 and rescaled to the bounding box of the displayed cortical surface for visualization. The viewer can display the resulting streamlines alongside cortical surfaces and EEG electrode locations. Streamline subsampling is used to reduce rendering demands and affects the visible density of fibres; it is not used as a quantitative measure of tract strength, prevalence, or anatomical correspondence.

\subsection*{Resource organization and reproducibility}

The repository includes the main viewer, a standalone effective-connectivity implementation, an atlas manifest, scripts for preparing cortical surfaces and retrieving tractography, and command-line utilities for EEG quality control, preprocessing, connectivity, and LORETA-related outputs. The connectivity implementation can be supplied by the local \texttt{effconn.py} module or by a compatible installation of \texttt{effconnpy}. The project is distributed under Apache License 2.0, while external atlases, templates, and tractography resources remain subject to their original licenses and citation requirements.

For reproducibility, the software version or commit, Python and dependency versions, input data provenance, external resource versions, and analysis settings should be archived. The resource manifest should be checked against the actual files. Quality-control reports should distinguish between a metadata entry and an asset that exists, loads successfully, and renders correctly.

\section*{Acknowledgements}
The authors would like to thank the Open communities of Neurostars for their feedback.

\section*{Funding}
The research presented in this article received partial funding from the Polish Ministry of Science and Higher Education assigned to the AGH University of Science and Technology in Krakow.

\section*{Author contributions}

A.C. conceived and coordinated the project, developed the software, performed the validation, and wrote the original manuscript. P.G. contributed to software development, scientific validation, and manuscript review.

\section*{Competing interests}

The authors declare no competing interests.

\bibliographystyle{unsrtnat}
\bibliography{sample}

\end{document}